# Planar Ultrananocrystalline Diamond Field Emitter in Accelerator RF Electron Injector: Performance Metrics

Sergey V. Baryshev,[1,2*] Sergey Antipov,[1,2] Jiahang Shao,[2†] Chunguang Jing,[1,2] Kenneth J. Pérez Quintero,[3‡], Jiaqi Qui,[1,2] Wanming Liu,[2] Wei Gai,[2] Alexei D. Kanareykin,[1] and Anirudha V. Sumant[3§]

[1]Euclid TechLabs, 365 Remington Blvd., Bolingbrook, IL 60440, USA
[2]High Energy Physics Division, Argonne National Laboratory, 9700 S. Cass Ave., Argonne, IL 60439, USA
[3]Center for Nanoscale Materials, Argonne National Laboratory, 9700 S. Cass Ave., Argonne, IL 60439, USA

A case performance study of a planar field emission cathode (FEC) based on nitrogen-incorporated ultrananocrystalline diamond, (N)UNCD, was carried out in an RF 1.3 GHz electron gun. The FEC was a 100 nm (N)UNCD film grown on a 20 mm diameter stainless steel disk with a Mo buffer layer. At surface gradients 45-65 MV/m, peak currents of 1-80 mA (equivalent to 0.3-25 mA/cm$^2$) were achieved. Imaging with two YAG screens confirmed emission from the (N)UNCD surface with (1) the beam emittance of 1.5 mm×mrad/mm-rms, and (2) longitudinal FWHM and rms energy spread of 0.7% and 11% at an electron energy of 2 MeV. Current stability was tested over the course of $36\times10^3$ RF pulses (equivalent to $288\times10^6$ GHz oscillations).

Despite being successfully deployed for RF microelectronics for decades[1] and considered conceptually for accelerator applications,[2,3] it is only now that field emitters are making inroads into the applied area of scientific and industrial accelerators driven by RF guns.[4-6] Field emitters are generally shaped in the form of a wire/cone/pyramid with a sharp tip of a few tens of nm in order to locally enhance electric fields to the GV/m range under modest applied macroscopic fields of the order of 1-10 MV/m. When directly subjected to an electric RF field in the injector, the FEC may significantly simplify the architecture of RF electron guns. In a straightforward picture, electron bunches are generated and phased by the RF electric field itself every time a positive electric field peaks on the FEC's surface, and high repetition rates equal to the RF frequency are supported automatically. A simplified injector on an FEC platform would greatly benefit superconducting RF (SRF) technology by leading to fully cryogenic compact SRF linacs with 50-60% wall-plug efficiency for medical and industry applications. SRF examples include electron-linac factories for Mo-99 production for nuclear medicine to rule out weapons grade uranium from the production cycle or compact bright inverse Compton sources for basic science and semiconductor lithography. These and many other applications (like cargo inspection) require normal conducting or SRF systems delivering beam power of 10 to 100 kW at electron energy 10 to 50 MeV. Thus, average currents of 1 to 10 mA are needed. Operation of high-duty-cycle ($10^{-3}$ to CW) accelerators with FEC as a source would compel the use of advanced materials for its design so that the FEC performs stably under high repetition rate or CW conditions inducing high electric and/or thermal loads. Synthetic diamond thin films and structures are among the most promising materials for high-frequency field emission applications which include both vacuum microelectronics[7] and accelerators.[5] In Ref.[5], currents of ≥10 mA were produced by directly subjecting a diamond tip array to an RF field in an L-band injector. With RF gradients of about 25 MV/m, the local actuating electric field on a tip was as high as ~1 GV/m. This example with an electron source of a simple design indeed suggests that this route to a compact RF gun architecture is feasible in practice. But can it be simplified even further while preserving or improving overall performance of the electron source?

Herein a case study of the performance of a planar FEC based on a nitrogen-incorporated ultrananocrystalline diamond, (N)UNCD, thin film is presented. The FEC was conditioned to 66 MV/m in an RF L-band gun operated at 1.3 GHz. An imaging system with YAG screens downstream the RF injector and a

---

* sergey.v.baryshev@gmail.com
† Also with Department of Engineering Physics, Tsinghua University, Beijing 100084, People's Republic of China
‡ Also with Physics Department, University of Puerto Rico, Río Piedras Campus, San Juan, PR 00931, USA
§ sumant@anl.gov

Faraday cup was used to quantitatively assess and characterize electron emission from the planar (N)UNCD FEC. Our choice arises from the following facts. First, planar as-grown (N)UNCD films have been shown to be an excellent field emitter with a turn-on DC voltage as low as 1.5 V/μm and current stability over 1,000 hours.[8,9] Hence, it may be of use in both normal and SRF linac systems. Second, previous experiments on nanocrystalline (NCD)[10,11] and UNCD[11,12] suggest that the electron field emission originates from grain boundaries (GBs). Hence, the total current is proportional to the fractional area of GBs in the UNCD film; and UNCD should have the highest area among other synthetic polycrystalline diamonds. Third, compared to the Spindt field emiter[1] with ~$10^8$ tips/cm$^2$ UNCD has ~$10^{13}$ emitting GBs per cm$^2$. Hence, the current load per emitter is reduced by a few orders of magnitude, when producing similar total macroscopic current. Four, planar UNCD production avoids any lithography and transfer steps, and can be scaled from a few millimeters to a wafer-scale process (150 mm-200 mm diameter). Hence, an FEC based on UNCD may become a true commodity electron source. Five, for very high gradient (~>100 MV/m) applications, the planar UNCD FEC might be the method-of-choice. Unlike field emitters based on sharp features that must be placed in a relatively weak macroscopic field, one also might expect an excellent beam emittance from the planar FEC due to space charge mitigation at 100 MV/m or higher.

To synthesize an (N)UNCD film directly on a default cathode plug, the plug was made sectional with a detachable top stainless steel (SS) disk 3 mm in thickness and 20 mm in diameter. As a first step to ensure good nucleation of UNCD, a buffer molybdenum layer was deposited using a custom magnetron sputtering system with a base pressure ~$10^{-7}$ Torr. The Mo layer had roughly a thickness of 100-200 nm. (N)UNCD growth was identical to the process described earlier.[13] Fig.1a shows the final assembly of the cathode plug as it gets inserted into an RF electron gun. The SS disk with the (N)UNCD film on top is bolted to an aluminum cylinder base. Fig.1b represents visible Raman spectra, unique for (N)UNCD, recorded using a He-Ne laser (λ=633 nm).[14,15] In Fig.1b, spectra before and after high power tests with more than a billion RF bursts (see following sections for details) are compared confirming the high resistance of UNCD to external harsh high power conditions. Scanning electron microscopy (SEM) images in Fig.1c also confirm this conclusion: a uniform needle-like nanostructure, typical for (N)UNCD, was observed before and after high power testing.

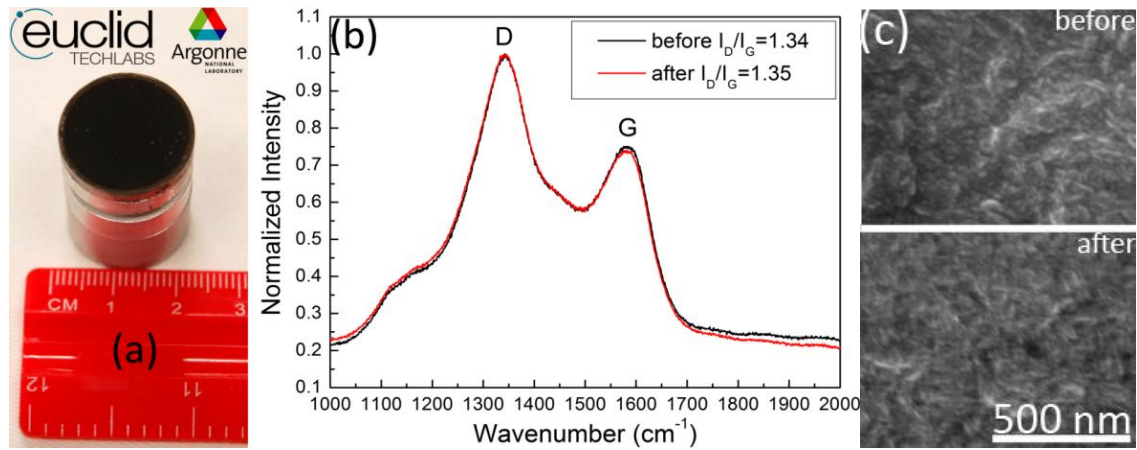

Fig.1. (a) A photograph of the cathode plug with (N)UNCD film deposited. (b) Visible Raman spectra and (c) SEM images, typical for (N)UNCD films, on Mo/SS before and after high power testing.

Field emission performance of the (N)UNCD planar cathode was tested in a dedicated test stand (see Fig.2). The RF electron injector was a half-cell standing wave copper cavity with a designed peak field of 1 to 120 MV/m on the cathode surface, depending on the input RF power.[16] The frequency of the cavity was tuned to 1.3 GHz to match the klystron frequency. The cathode plug is on a retractable actuator and can be replaced via detaching a flange on the injector's back wall. The injector cavity with the (N)UNCD FEC was evacuated to a base pressure ~$10^{-9}$ Torr. The principal timing scheme is presented in the inset in Fig.2. Quasi-rectangular RF pulses 6 μs long contained 8,000 oscillations of the 1.3 GHz RF frequency. RF pulses were separated in time by intervals corresponding to klystron's repetition rate. When conditioning the cavity to high power, the klystron was run at a 10 Hz repetition rate. A repetition rate of 1 Hz was used for measurements. Fig.2 shows the equipment used to characterize the electron beam emitted from the (N)UNCD FEC. A Faraday cup (FC) assessed the current/charge produced by an RF pulse. Thus, the peak charge can be estimated by dividing the total charge in the RF pulse by the number of GHz oscillations in the RF pulse. An imaging system consisting of a solenoid, steering (trim) magnets, and YAG screens located downstream of the injector was used to project and manipulate the beam emitted from the (N)UNCD FEC.

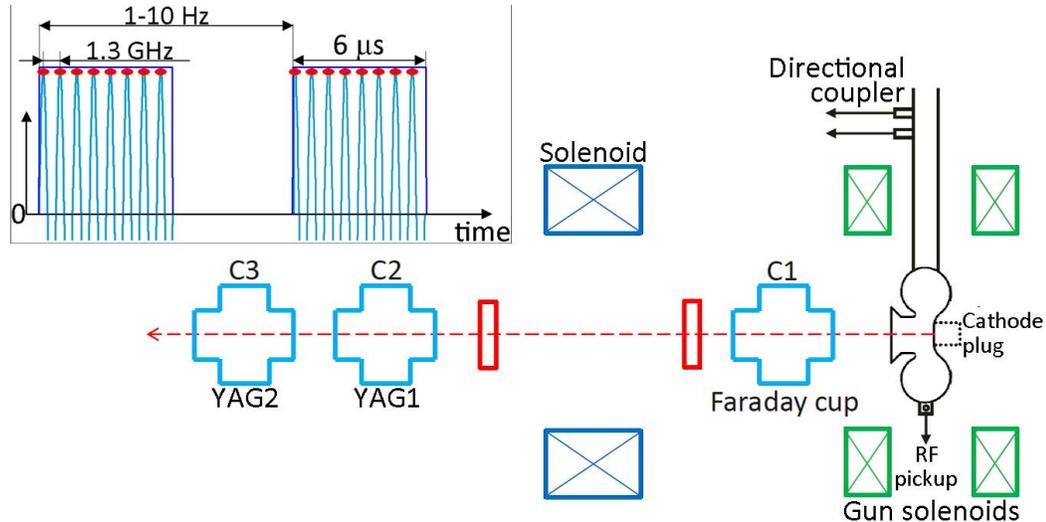

Fig.2. Diagram of the experimental setup. The inset shows principal timing scheme in FEC measurements (negative part of the GHz oscillations not shown); the red circles represent electron bunches.

The injector cavity was conditioned to sustain an RF power of 1.77 MW. Conditioning took about 15 hours at 10 Hz. This corresponded to ~$10^6$ RF pulses and ~$10^9$ electron bunches. Breakdown events were monitored by an X-ray photomultiplier tube, a mirror in the C1 chamber capturing visible-light flashes, and an RF pickup probe measuring transmitted/reflected RF power in the cavity. There were only 3 breakdowns detected on the (N)UNCD surface over the course of conditioning, while there were 100's of breakdowns on the cathode host bore edge. This result of a few breakdowns on the planar cathode surface is very typical. An identical situation holds when a polished copper photocathode is being conditioned. The significant difference here is that unlike a planar copper surface (no emission in Fig.3a), the (N)UNCD film was capable of producing an intense emission current comparable with that from the cathode bore edges.

To quantify the observed field emission from the (N)UNCD surface, we used a planar non-emitting Cu cathode as a reference.[17] In that study, the cavity frequency was tuned such that the cavity resonated at 1.3 GHz with the Cu cathode retracted 3 mm away from its zero position (i.e. flush with an inner back wall of the cavity). For the same reason, the (N)UNCD FEC was placed in the identical position. Since the curvature radius of the bore edge can be carefully defined, from electromagnetic simulations by SUPERFISH[18] it follows that 1.77 MW input power corresponded to electric fields of 65.5 and 212 MV/m on the cathode's surface and the cathode bore edge ring, respectively. This is why, for a correct direct comparison, the total charge collected by the FC versus RF power is plotted in Fig.3d for both Cu and (N)UNCD planar cathodes. It is assumed that, (1) the field enhancement factor and emitter area of the copper edge is the same in both cases, and (2) the planar Cu cathode case (Fig.3a) represented by blue solid circles in Fig.3d had only one electron emission component originating from the bore edge ring. Finally, using the data plotted in Fig.3d we subtract the parasitic bore ring dark current charge from the total charge recorded by the FC for the (N)UNCD FEC. The corrected (N)UNCD field emission data is plotted in Fig.3e. To make it generally relevant, the results are also plotted in terms of peak current (current per single GHz oscillation) versus electric field gradient. The 6 µs RF pulse contains $N_{GHz}=8\times10^3$ GHz oscillations at f=1.3 GHz. As commonly accepted,[3] we assume that emission takes place within 60° of phase for every GHz oscillation (±30° around the RF electric field maximum, i.e. 1/6 of the positive part of the GHz oscillation). Then peak current can be calculated as $I_{peak} = \frac{Q_{GHz}}{1/6 \times 1/f}$, where $Q_{GHz} = \frac{Q_{RF\,pulse}}{N_{GHz}}$. In Figs.3(d,e), the posted Fowler-Nordheim (F-N) factors, β, were extracted via linear approximation of the Q-E curves plotted in F-N coordinates, i.e. $\log\left(\frac{Q}{E^{5/2}}\right)$ versus $\frac{1}{E}$. The power 2.5 (instead traditional 2 in DC case) comes from a fact that the field emission needs to be time averaged in the RF case.[19]

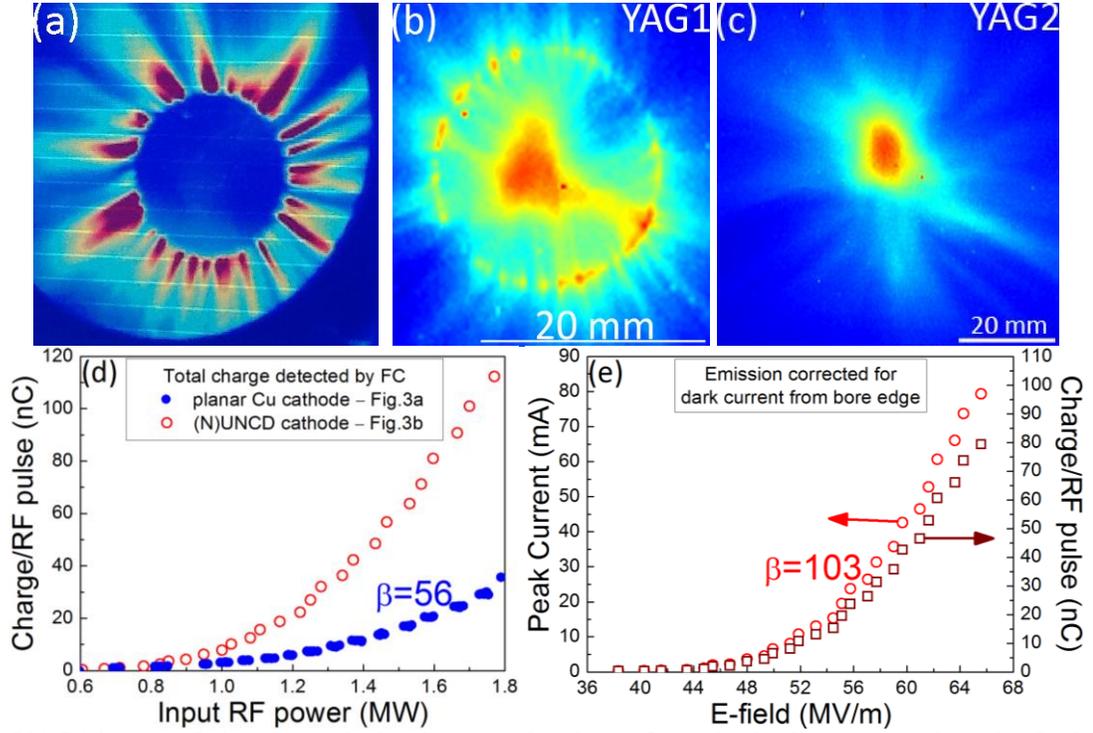

Fig.3. (a) YAG1 image of electron emission in case of a planar Cu cathode, 3 mm away from its flush position. The Cu cathode surface is the dark blue circle surrounded by red streaks. The streaks are wide-phase-spread dark current emission from the cathode bore edges. (b) YAG1 image of the electron beam from the planar (N)UNCD FEC, 3 mm away from its flush position. The beam is represented at its waist. (c) The same beam as in panel (b) projected on a downstream YAG2 screen. (d) Total charge per RF pulse as recorded by the FC in case (a) marked with blue solid circles, and in case (b) marked with red open circles. (e) Field emission characteristics (charge per RF pulse and peak current versus surface electric field) of the (N)UNCD FEC after the dark current from the Cu bore edge was subtracted.

In DC[20] and RF[17, 21] experiments, it was demonstrated for metals that the product of β and the maximum surface electric field, referred to as the maximum local electric field $E_{max}^{loc}$, is ~10 GV/m (1 volt per angstrom), and it is related to the fundamental electrical material strength. In the present case, the maximum power of 1.77 MW corresponded to $E_{max}^{loc}$ of 6.8 GV/m on the (N)UNCD FEC surface with β=103 and 11.8 GV/m on the Cu bore edge ring with β=56. This indicates that the achieved maximum power over the course of conditioning was not limited by the diamond cathode, but by the copper bore ring. The observation of 100's of breakdowns on the ring with only 3 occurring on the (N)UNCD FEC also confirms this conclusion. This suggests that with the measured β=103, the (N)UNCD FEC, if not limited by the bore edge ring, could be pushed to up to 96 MV/m with a peak current as high as 1 A. This further suggests that maximum gradient of 66 MV/m achieved corresponds to an operating point (peak current more than 10 times smaller than the projected 1 A at 96 MV/m) where the (N)UNCD FEC should perform stably. We crudely tested the stability of the electron emission from the diamond FEC. With the FC, two I-E curves (as in Fig.3e) were measured over a one hour period with the system running at 10 Hz repetition rate to improve statistics. 1 hour corresponded to $36\times10^3$ RF pulses and $288\times10^6$ GHz oscillations. Results are compiled in Table I. No significant deviation was observed.

**Table I. Stability test results**

| Time (s) / RF pulses / GHz oscillations | Peak current (mA) @ 45 MV/m | Peak current (mA) @ 55 MV/m | Peak current (mA) @ 65 MV/m |
|---|---|---|---|
| 0 s / 0 / 0 | 1.56±0.08 | 19.54±0.98 | 79.37±3.97 |
| 3,600 s / $36\times10^3$ / $288\times10^6$ | 1.47±0.07 | 19.24±0.96 | 79.26±3.96 |

Being limited by capabilities of the experimental beam line, we inspect the longitudinal energy spread of the emitted electrons semi-empirically instead of direct measurement. From SUPERFISH simulations, the energy of an electron at the exit of the injector was calculated as a function of the phase of the GHz oscillation. The experimentally determined $E_{max}^{loc}=\beta\times E_{gradient}$ of 6.8 GV/m was directly inserted into the F-N equation,[19] and allowed determination of the current versus the phase of the GHz oscillation (see Fig.4a). Energy and current dependences on the phase superimposed in Fig.4a resulted in the electron energy spectrum, which is presented in Fig.4b. The FWHM longitudinal energy spread was deduced as 14 keV and the rms longitudinal energy spread (with 85% electrons counted) was 220keV. This remarkable energy spread was indirectly confirmed by imaging the electron beam on YAG1 while scanning with the steering magnets. The beam moved as a whole, preserving its shape. The obtained energy spread is case-specific, and can be improved simply by cavity design. Following Ref.[6], an injector can be made two-sectional with a varied cathode cell length. Optimization of this parameter should lead to the rms energy spread as low as 1%. It means the electron energy spectrum will shape into a narrow single-peak curve because of the best overlapping between the phase dependences of energy and current (the energy-phase curve will change as cavity geometry is varied, and will not necessarily look like in Fig.4a).

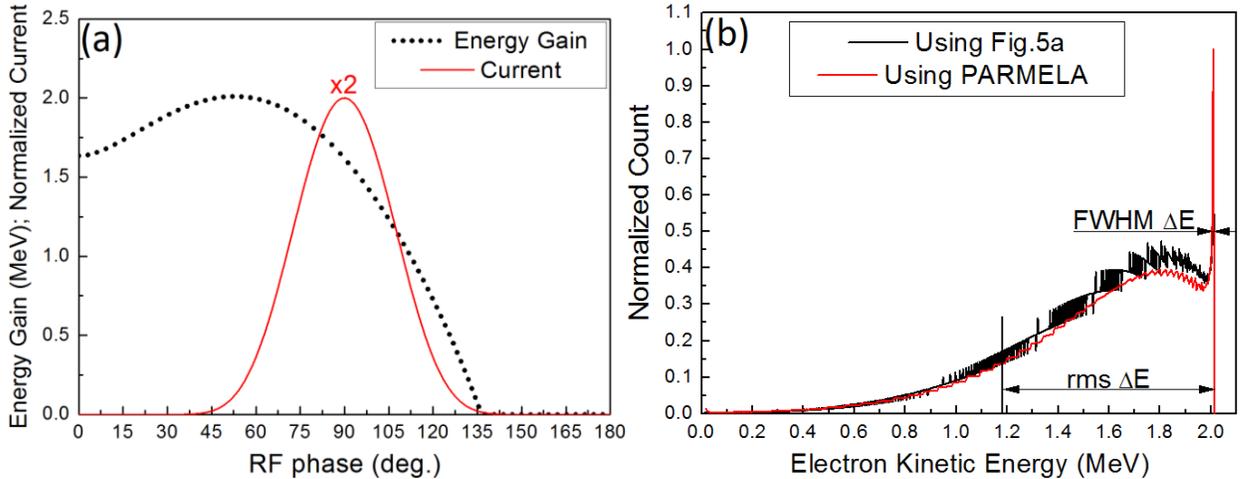

Fig.4. Energy spread calculation: (a) phase dependence of the energy gain of an electron in the cavity superimposed on the current directly calculated from the F-N equation with the experimentally measured $\beta\times E_{gradient}$=6.8 GV/m; (b) electron energy spectrum computed semi-empirically using the results from panel (a) and the electron tracking code PARMELA.

As the energy spread is narrow, the two beam projections presented in Fig.3b and Fig.3c can be used to deduce the emittance of the generated electron beam by the photocathode formula:

$$\varepsilon = \frac{\sigma_{waist}}{L}\times\sqrt{\sigma_L^2-\sigma_{waist}^2}\ , \quad (1)$$

where $\sigma_{waist}$ is the radius of the beam waist, and $\sigma_L$ is the beam radius projected on a second screen placed at a distance L. The sign σ refers to the standard deviation measure of Gaussian beams. From the YAG1 (Fig.3b) and YAG2 (Fig.3c) images, $\sigma_{waist}$ and $\sigma_L$ were measured as 1.87 mm and 4.20 mm, respectively, being averaged over orthogonal *x* and *y* directions. Given the distance L 465 mm and the (N)UNCD radius 10 mm, the beam emittance value was 15 mm×mrad, or equivalently 1.5 mm×mrad/mm-rms when normalized to the emitting area radius.

In conclusion, UNCD may serve as an excellent platform to create high efficiency and stable field emitters for a variety of high and low gradient accelerator applications. It can be of a planar or shaped geometry and assembled as a diode or triode, or can be operated in a single-electrode configuration directly subject to a strong electric field on its surface. Using the simplest single-electrode configuration, a case study of the thin-film conductive (N)UNCD as a FEC was performed. Field emission originating from the planar (N)UNCD surface was confirmed. Key metrics of the FEC were determined. At gradients 45 to 65 MV/m, peak currents 1 to 80 mA (current densities 0.3 to 25 mA/cm$^2$) were demonstrated. At the nominal electron energy of 2 MeV at the injector exit, the longitudinal FWHM and rms energy spreads were 0.7% and 11%, respectively. The beam emittance was 1.5 mm×mrad/mm-rms. Current stability was tested over the course of 1 hour at a repetition rate of 10 Hz, corresponding to $36×10^3$ RF pulses or to $288×10^6$ GHz oscillations. Upon finishing the high power testing, Raman spectroscopy showed no evidence of change in the intrinsic chemical phase of the (N)UNCD confirming its robustness.

**Acknowledgement**

The authors thank Vyacheslav Yakovlev (Fermilab) for motivating discussions. The work at the Center for Nanoscale Materials, a U.S. Department of Energy, Office of Science, Office of Basic Energy Sciences User Facility was performed under Contract No. DE-AC02-06CH11357. The work at the Argonne Wakefield Accelerator Facility was funded through the U.S. Department of Energy Office of Science under Contract No. DE-AC02-06CH11357. The electron microscopy was accomplished at the Electron Microscopy Center at Argonne National Laboratory, a U.S. Department of Energy Office of Science Laboratory operated under Contract No. DE-AC02-06CH11357 by UChicago Argonne, LLC. Funding was provided, in part, by NASA EPSCoR (grant No.NNX13AB22A) and NASA Space Grant (grant No. NNX10AM80H).